\documentclass[%
 reprint,
superscriptaddress,
 amsmath,amssymb,
 aps,
pra,
]{revtex4-2}
\usepackage[utf8]{inputenc}

\usepackage{graphicx}%
\usepackage{dcolumn}%
\usepackage{orcidlink}
\usepackage{bm}%
\usepackage{hyperref}%
\usepackage{commath}
\usepackage{mathtools}
\usepackage{amsmath}
\usepackage{xcolor}

\begin{document}

\preprint{APS/123-QED}

\title{Trapping of electrons and \textsuperscript{40}Ca\textsuperscript{+} ions in a dual-frequency Paul trap}

\author{Vladimir Mikhailovskii\,\orcidlink{0000-0003-1168-771X}}
\thanks{These authors contributed equally.}
\affiliation{Helmholtz-Institut Mainz, 55128 Mainz, Germany}
\affiliation{GSI Helmholtzzentrum f{\"u}r Schwerionenforschung GmbH, 64291 Darmstadt, Germany}
\affiliation{QUANTUM, Institut für Physik, Johannes Gutenberg-Universit{\"a}t, 55128, Mainz, Germany}

\author{Natalija Sheth}
\thanks{These authors contributed equally.}
\affiliation{Helmholtz-Institut Mainz, 55128 Mainz, Germany}
\affiliation{GSI Helmholtzzentrum f{\"u}r Schwerionenforschung GmbH, 64291 Darmstadt, Germany}
\affiliation{QUANTUM, Institut für Physik, Johannes Gutenberg-Universit{\"a}t, 55128, Mainz, Germany}

\author{Guofeng Qu\,\orcidlink{0000-0001-8840-3004}}
\email{Corresponding author:quguofeng@scu.edu.cn}
\affiliation{Institute of Nuclear Science and Technology, Sichuan University, 610065, Chengdu, China}
\affiliation{Key Laboratory of Radiation Physics and Technology of the Ministry of Education, Institute of Nuclear Science and Technology, Sichuan University, 610064 Chengdu, China}

\author{Michal Hejduk\,\orcidlink{0000-0002-4417-4817}}
\affiliation{Charles University, Faculty of Mathematics and Physics, Prague 8, Czech Republic}

\author{Niklas Vilhelm Lausti\,\orcidlink{0000-0001-9906-6971}}
\affiliation{Charles University, Faculty of Mathematics and Physics, Prague 8, Czech Republic}

\author{K. T. Satyajith\,\orcidlink{0000-0002-3794-035X}}
\affiliation{Delta Q, IMJ Institute of Research
\& Department of Physics,
Moodlakatte Institute of Technology, 576217, Moodlakatte, Karnataka, India}

\author{Christian Smorra\,\orcidlink{0000-0001-5584-7960}}
\affiliation{QUANTUM, Institut für Physik, Johannes Gutenberg-Universit{\"a}t, 55128, Mainz, Germany}
\affiliation{Heinrich Heine University Düsseldorf, 40225 Düsseldorf, Germany}

\author{Günther Werth}
\affiliation{QUANTUM, Institut für Physik, Johannes Gutenberg-Universit{\"a}t, 55128, Mainz, Germany}

\author{Neha Yadav}
\affiliation{Department of Physics, University of California, 94720-7300, Berkeley, USA}

\author{Qian Yu\,\orcidlink{0000-0002-8331-9417}}
\affiliation{Department of Physics, University of California, 94720-7300, Berkeley, USA}

\author{Clemens Matthiesen\,\orcidlink{0000-0001-7842-6536}}
\thanks{Current address: Oxford Ionics, Oxford, OX5 1PF, UK}
\affiliation{Department of Physics, University of California, 94720-7300, Berkeley, USA}

\author{Hartmut Häffner\,\orcidlink{0000-0002-5113-9622}}
\affiliation{Department of Physics, University of California, 94720-7300, Berkeley, USA}

\author{Ferdinand Schmidt-Kaler\,\orcidlink{0000-0002-5697-2568}}
\affiliation{QUANTUM, Institut für Physik, Johannes Gutenberg-Universit{\"a}t, 55128, Mainz, Germany}
 
\author{Hendrik Bekker\,\orcidlink{0000-0002-6535-696X}}
\affiliation{Helmholtz-Institut Mainz, 55128 Mainz, Germany}
\affiliation{GSI Helmholtzzentrum f{\"u}r Schwerionenforschung GmbH, 64291 Darmstadt, Germany}
\affiliation{QUANTUM, Institut für Physik, Johannes Gutenberg-Universit{\"a}t, 55128, Mainz, Germany}

\author{Dmitry Budker\,\orcidlink{0000-0002-7356-4814}} 
\email{Corresponding author: budker@uni-mainz.de}
\affiliation{Helmholtz-Institut Mainz, 55128 Mainz, Germany}
\affiliation{GSI Helmholtzzentrum f{\"u}r Schwerionenforschung GmbH, 64291 Darmstadt, Germany}
\affiliation{QUANTUM, Institut für Physik, Johannes Gutenberg-Universit{\"a}t, 55128, Mainz, Germany}
\affiliation{Department of Physics, University of California, 94720-7300, Berkeley, USA}

\date{\today}%

\begin{abstract}

We demonstrate the operation of a dual-frequency Paul trap and characterize its performance by storing either electrons or calcium ions while applying two quadrupole fields simultaneously which oscillate at $\Omega_\textrm{fast} = 2\pi \times 1.6$\,GHz and $\Omega_\textrm{slow} = 2\pi \times 2$\,MHz. The particles are loaded and stored in the trap under various conditions followed by detection employing an electron multiplier tube. We find that tens of electrons or ions can be trapped for up to ten milliseconds and a small fraction remains trapped even after hundreds of milliseconds. During dual-frequency operation we find that while the number of trapped electrons rapidly decreases with increase of the $\Omega_\textrm{slow}$ field amplitude, the number of trapped ions shows no dependence on the $\Omega_\textrm{fast}$ field amplitude as supported by our extensive numerical simulations.  We aim to use a similar trap for synthesising antihydrogen from antiprotons and positrons. Accordingly, we discuss open challenges such as the co-trapping of oppositely charged species and particle trap duration.
\end{abstract}

\maketitle

\section{Introduction}

Radiofrequency Paul traps are one of the most common traps in the field of atomic and molecular physics due to their reliability and excellent optical access. However, their use in the field of antimatter research is limited due to, for example, challenges with injecting charged particles from external sources and laser cooling of fundamental particles. On the other hand, radiofrequency (RF) Paul traps have the major advantage of allowing for the storage of low energy, oppositely charged particles in the same volume and at high density, whereas other methods require dynamically merging two particle clouds which limits the interaction time~\cite{amorettiProductionDetectionCold2002, andresenTrappedAntihydrogen2010, ahmadiAntihydrogenAccumulationFundamental2017}. The resulting enhanced efficiency of antihydrogen production and other antimatter atoms and molecules from their constituents was already noted by Hans Dehmelt in 1995 and more recently studied by some of us~\cite{Dehmelt1995, Leefer2017}. In the present work, we demonstrate an experimental realization of a dual-frequency Paul trap for the future co-trapping of antiprotons and positrons. We note that a combined RF Paul trap and Penning trap for the production of antihydrogen was demonstrated in 1995 but it saw limited use~\cite{Walz1995,pahlCombinedTrapLaser2000}. In our AntiMatter-on-a-Chip (AMOC) project we aim to ultimately produce and study antihydrogen in a table-top experiment in our laboratory.

CERN is currently the only place where cold antiprotons can be obtained but the BASE-STEP~\cite{Smorra2023, leonhardtProtonTransportAntimatter2025} and PUMA collaborations \cite{Aumann2022} are developing methods to transport them to other facilities using transportable Penning traps. We envision that with these novel tools and techniques, high-precision antimatter research will become commonplace in laboratories around the world enabling a more diverse range of experiments and applications. Previously mentioned disadvantages of RF traps can then be eliminated using, for example, similar techniques to those used for sympathetic cooling of highly charged ions in a cryogenic RF trap and image current detection of electrons~\cite{leopoldCryogenicRadiofrequencyIon2019,taniguchiImageCurrentDetection2025}.

Antimatter research is at the frontier of the search for physics beyond the Standard Model because of one of the key mysteries of modern physics: the observed asymmetry between matter and antimatter in the known universe \cite{Canetti2012}. High-resolution spectroscopy of antihydrogen is a promising approach~\cite{Ahmadi2018, Ahmadi2020, Khabarova2023}. High-precision measurements and comparisons of fundamental properties of matter and antimatter are also used to address this challenge, where the mass~\cite{Hori2011,Gabrielse1999,Ulmer2015}, charge \cite{Hughes1992}, and magnetic moments~\cite{Dehmelt1987,Disciacca2013} are investigated. In the latter case, the BASE-STEP experiment is specifically developed to perform measurements on antiprotons away from the electromagnetically noisy environment at the antiproton decelerator which limits the achievable precision~\cite{Smorra2023}.

In the present pioneering study, we develop an RF trap for co-trapping electrons and atomic ions which act as analogues for positrons and antiprotons. At the current stage of AMOC we prefer to use these easily accessible particles because, to a large degree, only their charge-to-mass $q/m$ ratio affects the trap stability. In other words, the results for electrons can be applied to positrons simply by inverting the polarity of DC-electrodes. And instead of (anti)protons we use heavier $^{40}$Ca$^{+}$ ions coming from the same source as the electrons. Due to the very different $q/m$ ratio of electrons and ions, two frequencies have to be applied for stable trapping. Predominantly theoretical studies have been performed for this in the case of ions trapped in a dual-frequency field~\cite{Trypogeorgos2016,Foot2018,Snyder2016}. More recently, co-trapping of a nanoparticle with an atomic ion, where the $q/m$ differed by a factor of $10^6$, was experimentally demonstrated~\cite{Bykov2024}. In this work, we demonstrate the successful trapping of electrons, or much heavier Ca$^{+}$ ions in our dual-frequency RF trap and investigate the relevant properties of this device as a pathway to co-trapping these species.

Trapping electrons employing an RF trap is challenging due the need for GHz- instead of MHz-frequency fields and also the absence of laser cooling methods and laser induced fluorescence detection techniques. A few groups have managed to overcome these challenges~\cite{Walz1995, Matthiesen2021,taniguchiImageCurrentDetection2025} with a range of applications in mind, such as, quantum information and computation~\cite{Osada2022,Daniilidis2013, Kotler2017,Lausti2023}. Applying a second frequency to these devices requires careful tuning of the frequency and amplitude to ensure continued stable trapping of the electrons. Moreover, the application of such a second frequency poses practical challenges due to the common use of single-frequency resonators to enhance the trapping field.

This paper is organized as follows: Starting with the description of our experimental setup, we then develop the dual-RF trap theory applicable to our trap geometry to obtain predictions for the regions of stability. Subsequently, we describe the experimental results when loading and trapping electrons or $^{40}\textrm{Ca}^{+}$ ions while applying either one or two RF fields. We characterize the trapping time distribution and study trapping stability. Finally, we sketch future plans for improved trap designs.

\section{Experimental setup}
Here we describe the technical details of our setup, including the trap, particle loading method, and detection scheme. Depending on the trapped species and purpose of the measurement, different sets of RF, DC, and extraction voltages are applied.
\subsection{Trap design}

\begin{figure}[]
    \centering
    \includegraphics[width=\linewidth]{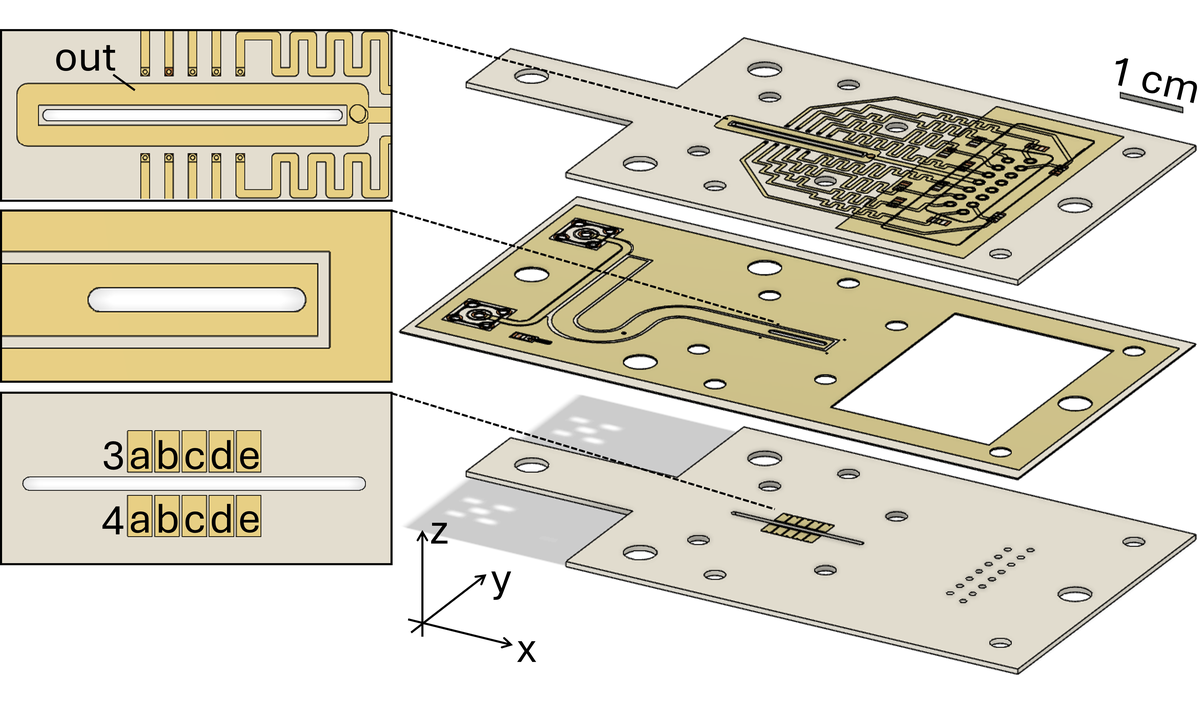}
    \caption{Schematic of the trap PCBs, based on \cite{Matthiesen2021}. The central one features a resonator to provide the quadrupole field in the slot shown in the middle inset. The identical top and bottom PCBs have ten DC electrodes on the side facing the central board as shown in the lower inset and an extraction electrode facing away from the center as shown in the upper inset.}
    \label{fig:trap}
\end{figure}

Our linear segmented Paul trap is assembled from three printed circuit boards (PCBs) which are separated by 1.27\,mm using ceramic spacers, see Fig.\,\ref{fig:trap}. The thickness of the central board is 0.2 mm and that of the others is 0.8 mm. The RF electrode consists of a capacitively coupled coplanar waveguide $\lambda/2$-resonator on the central PCB~\cite{Matthiesen2021}. It is shorted to ground at the center, has a resonance frequency of $\Omega_\textrm{fast} = 2\pi\times1.6$\,GHz, and a Q factor of 25. In a 9.0\,mm long and 0.9\,mm wide slot at the end of the resonator it generates a quadrupole field in the \textit{yz}-plane used to trap electrons. The top and bottom boards have 21.0\,mm long and 0.9\,mm wide slots parallel to the central slot that allow for optical access to the trap region. Those boards feature ten segmented DC electrodes symmetrically distributed on both sides of the slot on the PCB side facing the central board. Each block of five DC electrodes is labeled from \textit{a} to \textit{e} starting from the coupling side of the resonator. The outside surfaces of both PCBs also contain a single electrode around the slots used to extract particles.  The assembly is placed at the center of a vacuum chamber as shown in Fig.\,\ref{fig:setup}. The chamber is pumped by an Agilent TwissTorr 304 turbopump and a Saes NEXTorr Z 100 getter pump to a base pressure on the order of $10^{-10}$\,hPa.

Typically, an RF field with approximately 2\,W of power is applied at the resonator frequency $\Omega_\textrm{fast}$ to the in-coupling electrode, which results in the quadrupole field in the \textit{yz}-plane inside the slot of the resonator. To confine electrons in the \textit{x}-direction, the \textit{a} and \textit{e} DC electrodes are biased by $-6$\,V, the \textit{b} and \textit{d} ones by $-2$\,V, while the \textit{c} and extraction electrodes are kept at ground potential.

To trap ions, we apply a $\Omega_\textrm{slow}$ to one pair of diagonally opposite \textit{c} electrodes while keeping the other pair grounded. This generates another quadrupole field in the \textit{yz}-plane with its center coinciding with the field generated by the resonator. Confinement in the $x$-direction is again achieved using the remaining DC electrodes, in this case by applying positive voltages: the \textit{a} and \textit{e} electrodes are biased to $+5$\,V, the \textit{b} and \textit{d} are at $+3$\,V, and the extraction electrodes is grounded.

\subsection{Measurement sequence}
\label{sec:setup}

\begin{figure}[]
    \centering
    \includegraphics[width=\linewidth]{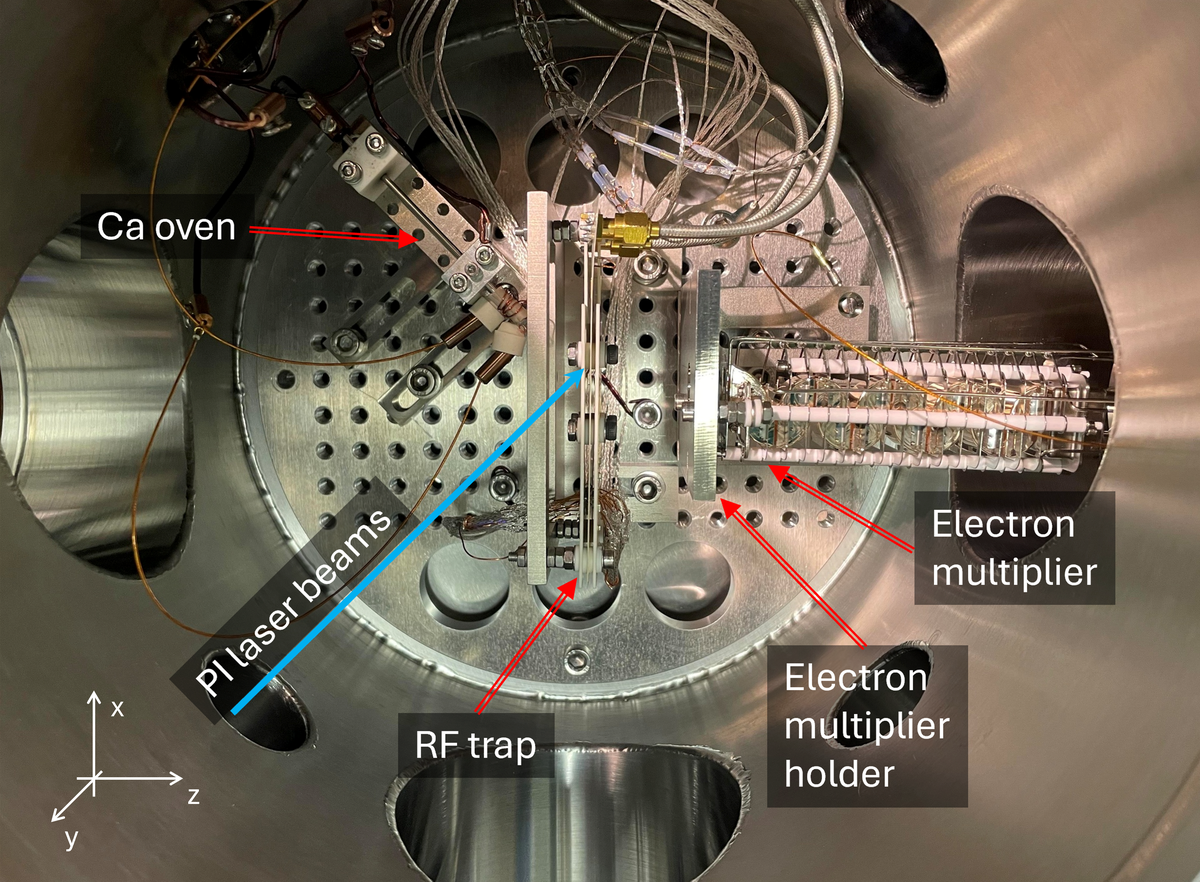}
    \caption{Top-down view of the experimental setup in the vacuum chamber. The trap PCBs, labeled RF trap, are seen from their side.\label{fig:setup}}
\end{figure}

\begin{figure}[]
    \centering
    \includegraphics[width=\linewidth]{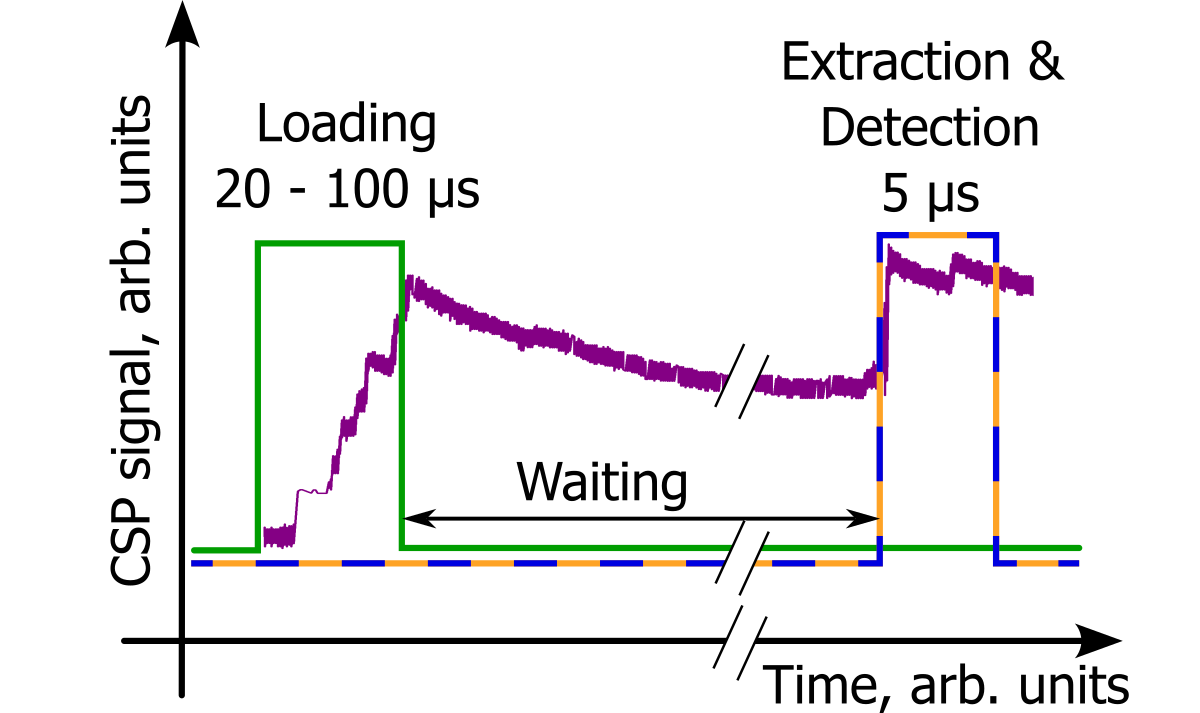}
    \caption{Timing sequence of a typical measurement cycle. Green line: Gate for the AOM driver of the photo-ionizing 390 nm laser beam. Orange line: extraction voltage. Blue line: detection trigger. Purple: signal from the charge-sensitive preamplifier in the case of electron detection. For ion detection the overall result is similar.\label{fig:sequence}}
\end{figure}

Most of the measurements presented here consist of three steps: loading of electrons or ions into the trap, storing them for a defined waiting time, and then extracting them for detection, as shown in Fig.\,~\ref{fig:sequence}. Electrons and ions are loaded into the trap by photo-ionization of calcium atoms. To this end, an atomic oven emits a Ca beam at a 45-degree angle with respect to the PCB surfaces so that it crosses the center of the trap. We employ the two-step photo-ionization scheme of $^{40}$Ca \cite{Gulde2001}, where a laser beam at 423\,nm is used to resonantly drive the $4^1S_0 - 4^1P_1$ transition in neutral Ca, and the second laser beam at 390\,nm is used for ionizing from the $4^1P_1$ state to near the continuum. This results in electrons and ions with sufficiently low energy for trapping. Both laser beams are coupled into the same single-mode fiber for good spatial alignment as confirmed by using a CCD camera to determine the beam properties. Subsequently, they are sent into the trap chamber through a viewport at a right angle to the calcium beam, see Fig.\,~\ref{fig:setup}. At the trap center, both laser beams are overlapping and have a power of approximately 1\,mW and the beam diameters are 20\,\textmu m and 80\,\textmu m for the laser beams at 390\,nm and 423\,nm, respectively.

During a measurement cycle, the atomic beam, 423\,nm laser beam, and RF field remain on. The 390\,nm beam is switched on only during the loading period by means of a single-pass acousto-optic modulator (AOM). Subsequently, the charged particles are stored in the trap for a defined waiting time which we vary from microseconds to maximally a second. This step ends by applying a $+1$\,V extraction pulse for 5\,\textmu s to the \textit{c} electrodes (in case of electrons) and $-1$\,V to \textit{b} and \textit{d} electrodes (in case of ions) on the side facing the detector, as well as to the extraction electrode on the same side. Here we use a Hamamatsu R596 electron multiplier tube (EMT) placed at a distance of 3\,cm from the trap. We apply $+140$\,V (in case of electrons) or $-500$\,V (for ions) to the front EMT holder to enhance collection efficiency without distorting potential inside the trap according to finite-element modeling. High voltage of 2000\,V is applied across the EMT resulting in a gain of approximately $10^{6}$.

The EMT output is amplified using a charge-sensitive preamplifier (CSP), model ORTEC 142PC, with a feedback capacitor of 0.1\,pF. The resulting signal is a voltage pulse with a nanoseconds rising edge and a slow decay of tens of microseconds; the amplitude of the pulses is proportional to the number of charges. Pulses are digitized using a CAEN DT5770 digitizer. Applying the $\Omega_\textrm{slow}$ field to the central electrodes for ion trapping introduces significant electronic pick-up in the detection system. To avoid this, we include a notch filter between CSP and digitizer with its central frequency at 2\,MHz. The digitizer is able to detect incoming pulses either continuously, using an internal trigger, or at defined time points, using an external trigger.

To calibrate the EMT and to optimize various parameters, such as the EMT holder potential, we use continuous detection. In this case, the RF field is off and the signal is due to electrons or ions (depending on the EMT potentials) generated by photo-ionization of calcium atoms. For trapping experiments when the RF field is on, the digitizer is triggered externally by the start signal of the detection window that coincides with the extraction of particles from the trap. During the loading time the EMT signal shows stepwise increases corresponding to overlapping fast-rise pulses at the output of the CSP, as shown in Fig.~\ref{fig:sequence}. These pulses are caused by charged particles that are not trapped and leave this region soon after production. When the extraction voltage is applied we observe a sharp increase in the signal. In comparison to the stepwise voltage increases during loading, this appears as a single pulse, suggesting that the detected particles arrive as a bunch due to their simultaneous extraction from the trap. Such behavior is observed only when all necessary components -- the RF field, DC voltages, and properly aligned laser beams -- are applied together. The appearance of this signal and its time correlation with the measurement cycle serve as primary evidence of successful particle trapping.

For measuring the amount of trapped particles, we repeat the sequence $10^{3}-10^{4}$ times and average over CSP pulses detected by the digitizer during the extraction pulses. Then the average height of the CSP pulses is converted to the average number of particles using the calibration procedure described in the Appendix \ref{sec:emt-calibration}. The uncertainty in the average number of particles is estimated from the CSP pulse amplitude histogram.

\section{Theory of dual-frequency Paul traps}
We briefly review the theory of general dual-frequency Paul traps and subsequently apply this to our trap geometry.
\subsection{General dual-frequency trap}
\label{sec:general}
The amplitude, $U$, and angular frequency, $\Omega$, of a field necessary for trapping charged particles is determined by their charge-to-mass ratio, $q/m$ \cite{Major2005}. For co-trapping species where this ratio differs roughly by an order of magnitude or more, the application of two fields is necessary. To ensure that both particle clouds are approximately of the same size, the voltages $U_\textrm{slow}$, $U_\textrm{fast}$ and frequencies $\Omega_{\textrm{slow}}$, $\Omega_{\textrm{fast}}$ of both fields should be chosen in such a way that the spring constants, $\kappa$, of both species in the oscillating potential are nearly equal \cite{Trypogeorgos2016}, meaning that
\begin{equation}
    \frac{\kappa_\textrm{e}}{\kappa_\textrm{Ca}} = \left(\frac{U_\textrm{fast}}{U_\textrm{slow}}\right)^2 \cdot \left(\frac{\Omega_\textrm{slow}}{\Omega_\textrm{fast}}\right)^2 \cdot \left(\frac{m_\textrm{Ca}}{m_\textrm{e}}\right) \approx 1\,.
\end{equation} 
Since the mass ratio $m_\textrm{Ca}/m_\textrm{e}$ is about 73\,000, this condition suggests that a frequency ratio $\eta := \Omega_\textrm{fast}/\Omega_\textrm{slow}$ on the order of $10^2$ is required. This mass difference implies that the electron motion is much faster than the ion motion, meaning that the higher frequency field $\Omega_\textrm{fast}$ is required to form the quadrupole potential for the electron trap while the slower moving ions are confined by the lower frequency field $\Omega_\textrm{slow}$. In our measurements and modeling we fix the frequency ratio to $\eta = \Omega_\textrm{fast} /\Omega_\textrm{slow} = 800$ as determined by the fixed resonator frequency and that of the low frequency notch filter. Experimentally we probe a range of Mathieu $q$ parameters by adjusting the amplitudes $U_\textrm{slow}$ and $U_\textrm{fast}$.

The stability of particles in dual-frequency traps has been studied in other works, for example~\cite{Leefer2017, Foot2018}. The time-dependent potential due to two RF fields in a quadrupole geometry can be written as
\begin{equation}
    \begin{split}
       \Phi(t, y, z)=\bigl[&U_0+U_\textrm{slow}\cos(\Omega_\textrm{slow} t)\\&+U_\textrm{fast}\cos(\Omega_\textrm{fast} t)\bigl]\frac{z^2-y^2}{2r_0^2}\,, 
    \end{split}
    \label{eq:pot}
\end{equation}
where $U_0$ is a DC potential and $r_0$ is the half-distance between opposite electrodes. By introducing the dimensionless time $\tau=\Omega_\textrm{fast}t/2$, the condition $\mathbf{F} = -\,e\,\boldsymbol{\nabla}\Phi$ results in the dimensionless Mathieu equations of motion,
\begin{equation}
    \begin{split}
      \overset{..}{u}\,\left(\tau\right) + \bigl[a_\textrm{u}& - 2q_\textrm{1}^u\cos(2\eta^{-1}\tau)\\&-2q_\textrm{2}^u\cos(2\tau)\bigl]u\,\left(\tau\right) = 0\,,  
    \end{split}
    \label{eq:gen-eqm}
\end{equation}
with $u=\{x, y\}$. Thereby, we defined 
\begin{equation}
    a_u=4\frac{eU_0}{mr_0^2\Omega_\textrm{fast}^2}\,,
\end{equation}
and
\begin{equation}
    q_\textrm{1,2}^z=-q_\textrm{1,2}^y=-2\frac{eU_\textrm{slow, fast}}{mr_{1,2}^2\Omega_\textrm{fast}^2}\,,
    \label{eq:q}
\end{equation}
with the mass $m$ of the trapped particle.

\begin{figure}
    \centering
    \includegraphics[width=\linewidth]{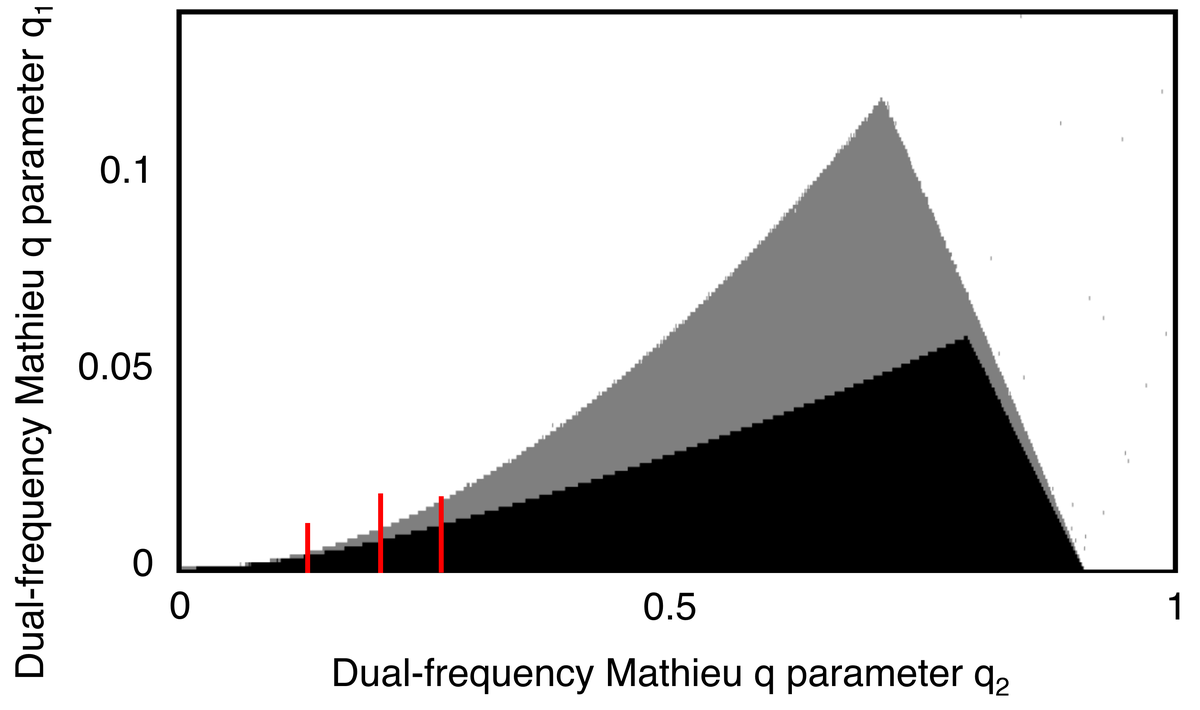}
    \caption{Stability diagram for $\eta=800$. The gray region represents stable sets of $\left(q_1, q_2\right)$ for general dual-frequency traps in $y$-direction. Along the $z$-direction, the signs of $q_1$ and $q_2$ flip. The black region shows the same for our trap geometry. The red lines indicate in which region we test the trapping of electrons experimentally. }
    \label{fig:stability}
\end{figure}

Because $\eta$ is a rational number, it is possible to use Floquet theory to find an analytical expression for the set of $q$'s and $a$'s that lead to stable trapping~\cite{Leefer2017}. For this, effectively, the zeros of a matrix determinant have to be found. A graphical representation of the stability region for $\eta = 800$ and $a_u=0$ is shown in Fig.\,\ref{fig:stability}. Note that this diagram is only for the stability along the $y$-axis.

In practice, it is difficult to apply two different frequencies to the same set of electrodes. This is mainly because they are often driven using resonators in order to reach sufficiently high field amplitudes, and these are challenging to implement for two-frequency operation. For example, the waveguide resonator of our trap strongly damps signals below 1.6\,GHz. Therefore, we use another set of electrodes to apply the low-frequency field. This has the added benefit of providing an opportunity to optimize their shape and distance $r_0$ for their respective frequency, which we will exploit in future trap versions. Moreover, using different sets of electrodes allows for an independent fine adjustment of the centers for both trapped species, if desired. In the following section, we describe the theory of our trap geometry in which two different sets of electrodes are operated at two different frequencies. 

\subsection{Stability of orbits}
\label{sec:ourtheory}

\begin{figure}
    \centering
    \includegraphics[width=\linewidth]{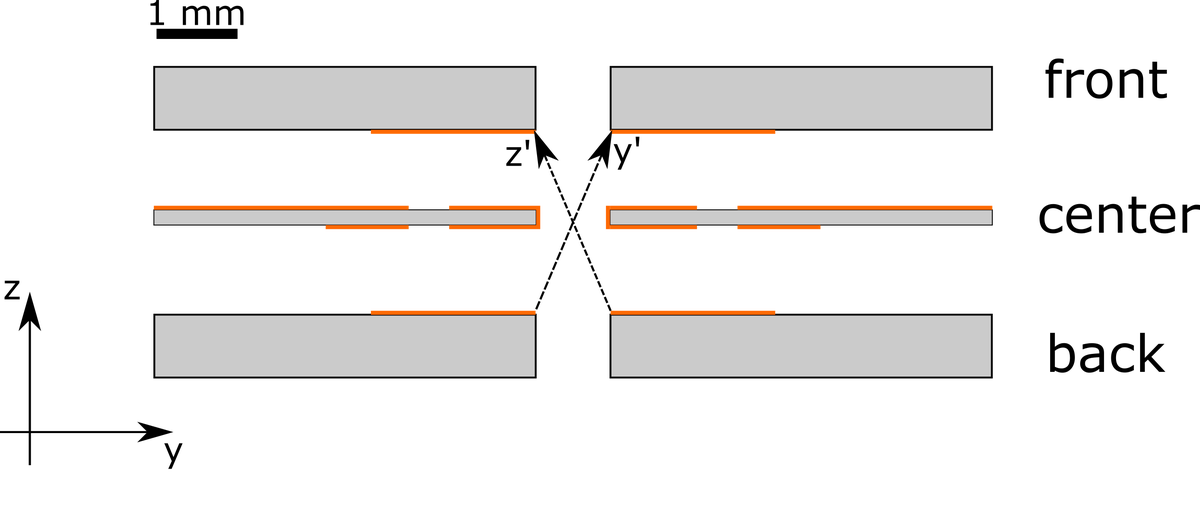}
    \caption{Cross section of the electron trapping region with the PCB substrate in gray and electrodes orange. Shown are the trap axes $y'$ and $z'$ relevant when $\Omega_\textrm{fast}$ is applied to the central DC electrodes for confining ions.}
    \label{fig:trap-b}
\end{figure}

While the central DC electrodes were not originally designed for RF operation and their geometry is not optimized for this purpose, the Paul trap demonstrates remarkable robustness. Even with flat and non-orthogonal pairs of electrodes, as shown in Fig.\,\ref{fig:trap-b}, we are able to trap ions. The resulting quadrupole field is in the same $yz$-plane as the field for confining electrons. To determine the equations of motion, the potential given in Eq.\,\eqref{eq:pot} has to be adapted to account for the different geometries of the electrodes. We associate the orthogonal $y$ and $z$-axes with the waveguide resonator to which we apply a signal of frequency $\Omega_\textrm{fast}$. On the other hand, the non-orthogonal $y'$ and $z'$-axes are associated with the DC electrodes to which a frequency $\Omega_\textrm{slow}$ is applied, see Fig.\,\ref{fig:trap-b}. The angle between the two axes $y$ and $y'$ is $\beta = 62$\,° so that we can make a coordinate transformation using the constants $c_1\coloneqq -\cos{2\beta}$ and $c_2\coloneqq \cos{4\beta}/\sin{2\beta}$. A particle in the trap then experiences the potential

\begin{equation}
    \begin{split}
        \Phi(t, y, z) & =U_\textrm{slow}\cos(\Omega_\textrm{slow} t)\frac{c_1y^2-c_1z^2+c_2yz}{2r_1^2}\\ & +U_{\textrm{fast, }y}\cos(\Omega_\textrm{fast} t)\frac{y^2}{2r_{2,y}^2}\\&-U_{\textrm{fast, }z}\cos(\Omega_\textrm{fast} t)\frac{z^2}{2r_{2,z}^2}\,,
    \end{split}
    \label{eq:pot-full}
\end{equation}
where we take into account the different dimensions $r_{2,y/z}$ and amplitudes $U_{\textrm{fast, }y/z}$ of the resonator in $y$- and $z$-directions, respectively.

This potential results in a coupled motion in $y$- and $z$-directions which is described by the equations of motion
\begin{subequations}
\begin{equation}
    \begin{split}
            \overset{..}{y}\left(\tau\right)&-\left[4c_1q_{1, y}\cos(2 \eta^{-1}\tau)+2q_{2, y}\cos(2\tau)\right]y\left(\tau\right)\\&+2c_2q_{1,z}\cos(2\eta^{-1}\tau)z\left(\tau\right) = 0\,,
    \end{split}
\end{equation}
\begin{equation}
    \begin{split}
            \overset{..}{z}\left(\tau\right)&-\left[4c_1q_{1,z}\cos(2 \eta^{-1}\tau)+2q_{2, z}\cos(2\tau)\right]z\left(\tau\right)\\&-2c_2q_{1,y}\cos(2\eta^{-1}\tau)y\left(\tau\right) = 0\,,
    \end{split}
\end{equation}
\label{eq:eqm}
\end{subequations}
with $\eta$ and $\tau$ as defined in Sec.\,\ref{sec:general}. The $q$-parameters are now given by
\begin{equation}
    q_{1,y}=-q_{1,z}=-2\frac{eU_{\textrm{slow}}}{mr_1^2\Omega_\textrm{fast}^2}\,,
    \label{eq:q1-our-geometry}
\end{equation}
and 
\begin{subequations}
    \begin{equation}
        q_{2,y}=-2\frac{eU_{\textrm{fast, }y}}{mr_{2,y}^2\Omega_\textrm{fast}^2}\,,
    \end{equation}
    \begin{equation}
        q_{2,z}=2\frac{eU_{\textrm{fast, },z}}{mr_{2,z}^2\Omega_\textrm{fast}^2}\,,
    \end{equation}
    \label{eq:q2-our-geometry}
\end{subequations}
where $m$ is the mass of the trapped particle species.

To solve the equations of motion, Eq.\,\eqref{eq:eqm}, for $u=\{y, z\}$ we employ the Runge-Kutta method of 4$^\textrm{th}$ order for various sets of $q$-parameters. Similar to the method in \cite{Leefer2017}, we introduce the stability parameter
\begin{equation}
    S = \frac{1}{\tau_2-\tau_1}\int_{\tau_1}^{\tau_2}u^2\left(\tau\right)\,\textrm{d}\tau\,,
\end{equation}
and consider a trajectory as stable only if $S < r_1^2$ when choosing $\tau_1 = 0$ and $\tau_2 = 100\,000$. These parameters ensure that the average amplitude of the particle motion does not exceed the trapping volume during 200\,000 periods of the $\Omega_\textrm{fast}$ field and 250 periods of the $\Omega_\textrm{slow}$ field.
With this condition, we can generate stability diagrams as shown in Fig.\,\ref{fig:stability}.

When comparing the stability region of the two different geometries of dual-frequency traps, we realize that our non-orthogonal electrode design puts more constraints on the low-frequency field. While we can find stable trajectories with values of $q_1$ up to 0.119 for general dual-frequency traps with orthogonal electrodes, the upper limit of $q_1$ for our trap is only 0.06 with our geometry.

In this work, we use six distinct Mathieu $q$ parameters. The trapping in a single-frequency RF trap is characterized by the parameter $q_e=2eU_\textrm{fast}/m_er_2^2\Omega_\textrm{1}^2$ for electrons and the parameter $q_{\textrm{Ca}^+}=2eU_\textrm{slow}/m_\textrm{Ca}r_1^2\Omega_\textrm{2}^2$ for calcium ions. In the context of dual-frequency trapping, the stability of electrons is defined by the two parameters $q_{1,\,e}$ and $q_{2,\,e}$, as defined in Eq.\,\eqref{eq:q1-our-geometry} and \eqref{eq:q2-our-geometry} by setting $m = m_e$. Setting $m = m_\textrm{Ca}$ results in the parameters $q_{1,\,\textrm{Ca}^+}$ and $q_{2,\,\textrm{Ca}^+}$ that describe the stability of calcium ions in the dual-frequency trap. In all cases, no distinction is made between the $y$- and $z$-direction, since it is assumed that the absolute values are equal along both axes.

\section{Experimental results and discussion}

As described in Sec.\,\ref{sec:setup} and supported by our theory predictions, we successfully trapped electrons and calcium ions separately. Next, we characterize the trap by studying the effects of dual-frequency operation on the motion of trapped particles and attempt to find the optimal working point.

\subsection{Motional frequencies of the electrons and ions}
\label{sec:e_trapping}

Application of a second RF field to the trapped particles can result in undesirable unstable conditions. Specifically, when the frequency of the second RF field is equal to the secular frequency $\omega$ of the particles macromotion, its amplitude is increased, resulting in particles loss. Since we keep Mathieu parameter $a_{y,z} = 0$ the secular frequency is related to the Mathieu $q$-parameter through $\omega = q \Omega / 2 \sqrt{2}$, and observation of this effect allows us to also compare the results with the theory developed in Sec.\,\ref{sec:ourtheory}.

\begin{figure}
    \centering
    \includegraphics[width=\linewidth]{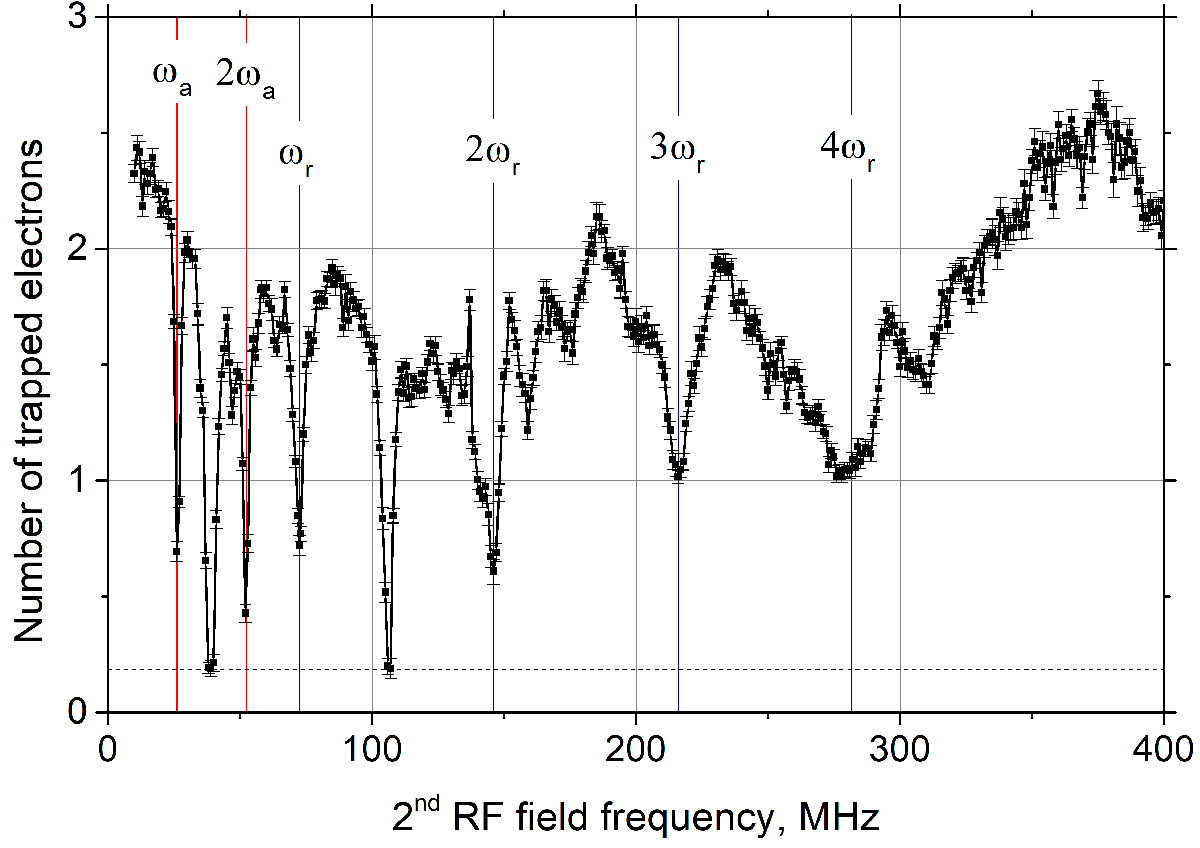}
    \caption{Dependence of the average number of trapped electrons on the second frequency, applied to the \textit{c} electrodes of the trap. The loading and waiting times were 100\,\textmu s and 500\,\textmu s, respectively. The dashed line indicates the background level.}
    \label{fig:MHz@resonator}
\end{figure}

To investigate this for electrons, we apply a second RF field to one pair of diagonally opposite $c$ electrodes, as it is done when trapping ions. We scan the frequency in the range from $2\pi\times10$ to $2\pi\times400$\,MHz in 1 MHz steps while measuring the amount of extracted electrons after a waiting time of 500\,\textmu s. The resulting spectrum shown in Fig.\,\ref{fig:MHz@resonator} features several dips which could be due to excitation of the macromotion. To confirm this and to identify the nature of the dips, we vary the DC potential and the RF drive power: resonances whose position shifts with DC bias are associated with axial motion while those sensitive to the power of $\Omega_\textrm{fast}$ field are associated with radial motion. The lowest axial frequency is measured at $\omega_a = 2\pi\times26$\,MHz and the lowest radial at $\omega_r = 2\pi\times72$\,MHz. Two resonances at $2\pi\times38$ and $2\pi\times106$\,MHz show no sensitivity to either DC or RF amplitude changes and coincide with significant MHz pickup in the detection chain and DC electrodes. We therefore suspect these two features to be spurious resonances of the wiring or electronics rather than motional frequencies of electrons.

\begin{figure}
    \centering
    \includegraphics[width=\linewidth]{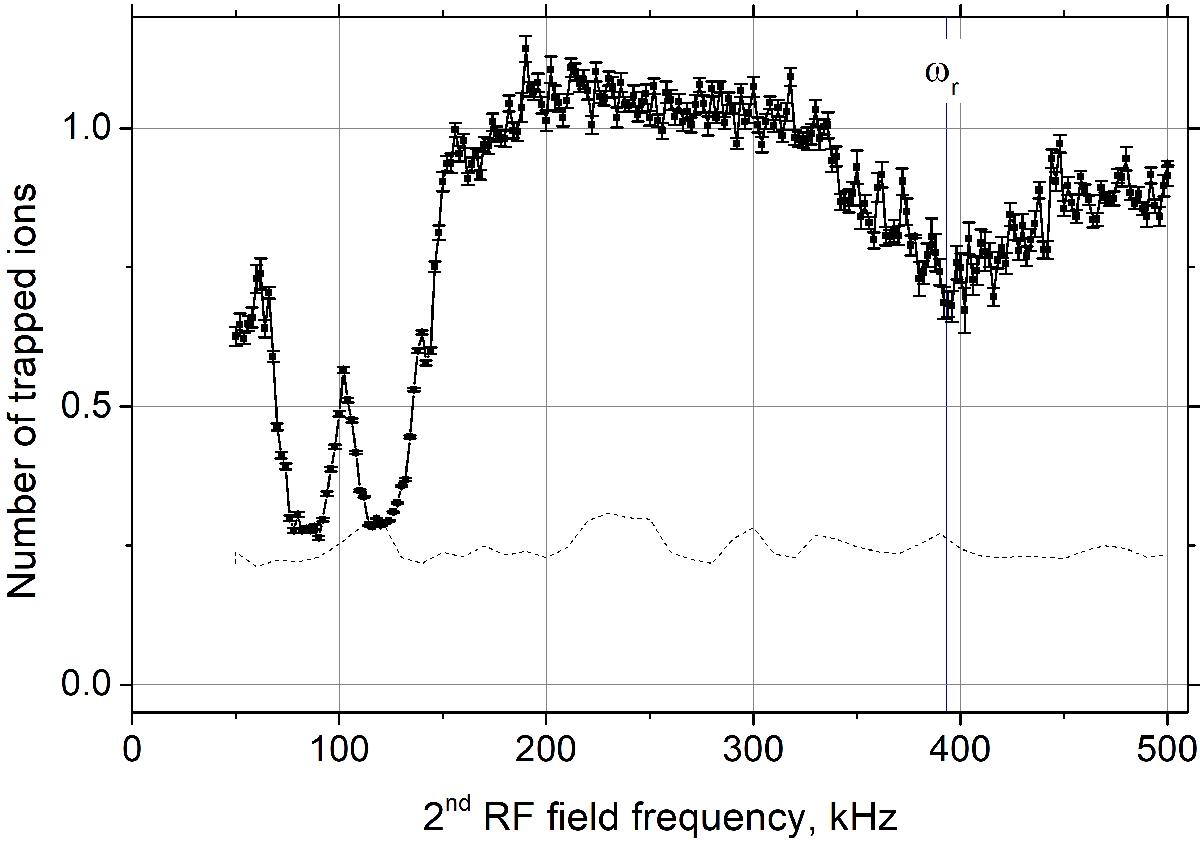}
    \caption{Average number of extracted Ca$^{+}$ ions as a function of the second frequency, applied to the \textit{d} electrode. Loading time: 100\,\textmu s, waiting time 1\,ms. The dashed line indicates the background level.}
    \label{fig:ionsVSf_khz}
\end{figure}

To study the frequencies of ion motion inside the trap, we apply an additional RF field to one of the \textit{d} electrodes of the trap. The amplitude of the corresponding voltage is 3\,V 
and we vary the frequency from $2\pi\times50$ to $2\pi\times500$\,kHz while measuring how many ions on average can be extracted after 1\,ms of waiting time. The results are shown in Fig.\,\ref{fig:ionsVSf_khz}. As the resonance at $2\pi\times395$\,kHz  shifts under variation of the MHz trapping voltage and has no sensitivity to DC variation, it is attributed to the radial motion of Ca$^{+}$ ions. 
Two dips at $2\pi\times80$\,kHz and $2\pi\times120$\,kHz are insensitive to variations of both DC electrode potentials and the $\Omega_\textrm{slow}$ voltage. Therefore, we again attribute them to spurious resonances not related to the secular motion of ions.

\subsection{Stability of particle orbits}

\begin{figure}
    \centering
    \includegraphics[width=\linewidth]{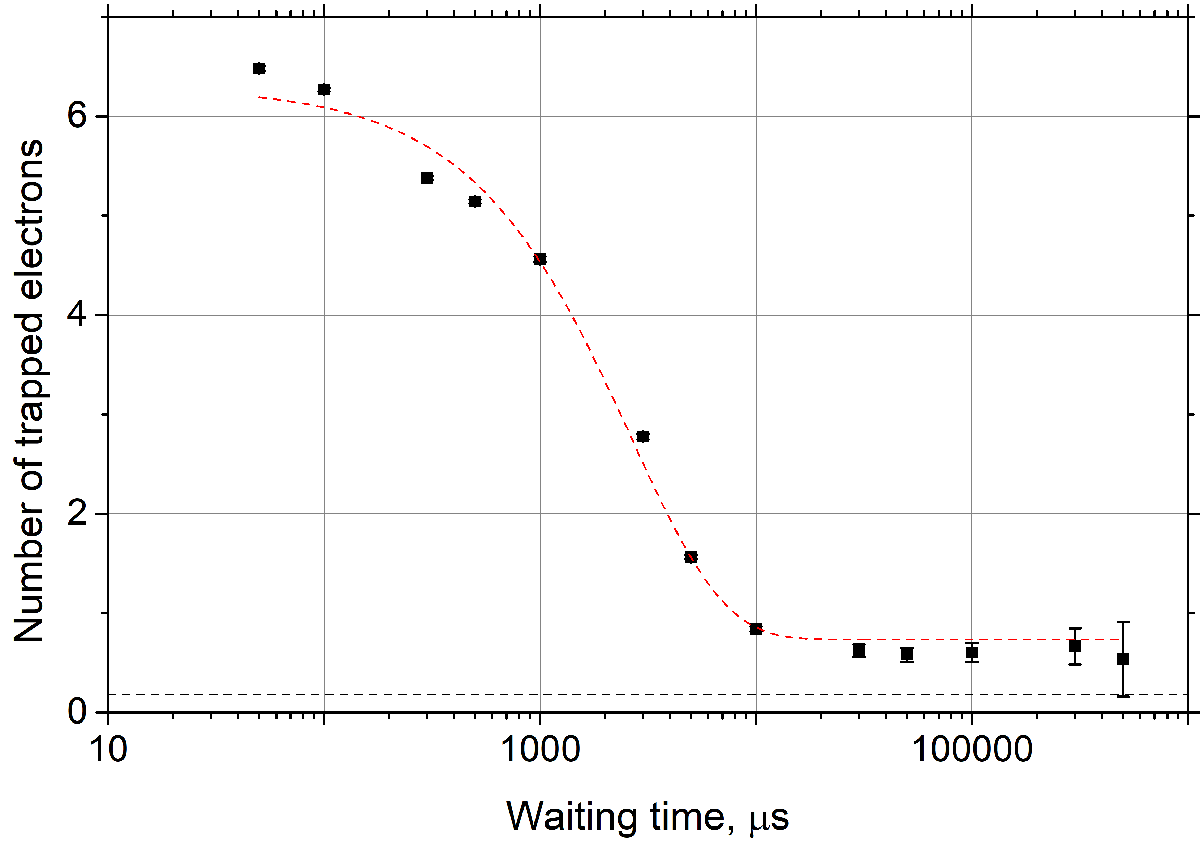}
    \caption{Average number of extracted electrons as a function of waiting time. Loading time: 50\,\textmu s. The red dashed line indicates the exponential decay model of Eq.~\ref{exp_decay}. The dashed line indicates the background level when the ionization lasers are blocked.}
    \label{fig:e-lifetime}
\end{figure}

\begin{figure}
    \centering
    \includegraphics[width=\linewidth]{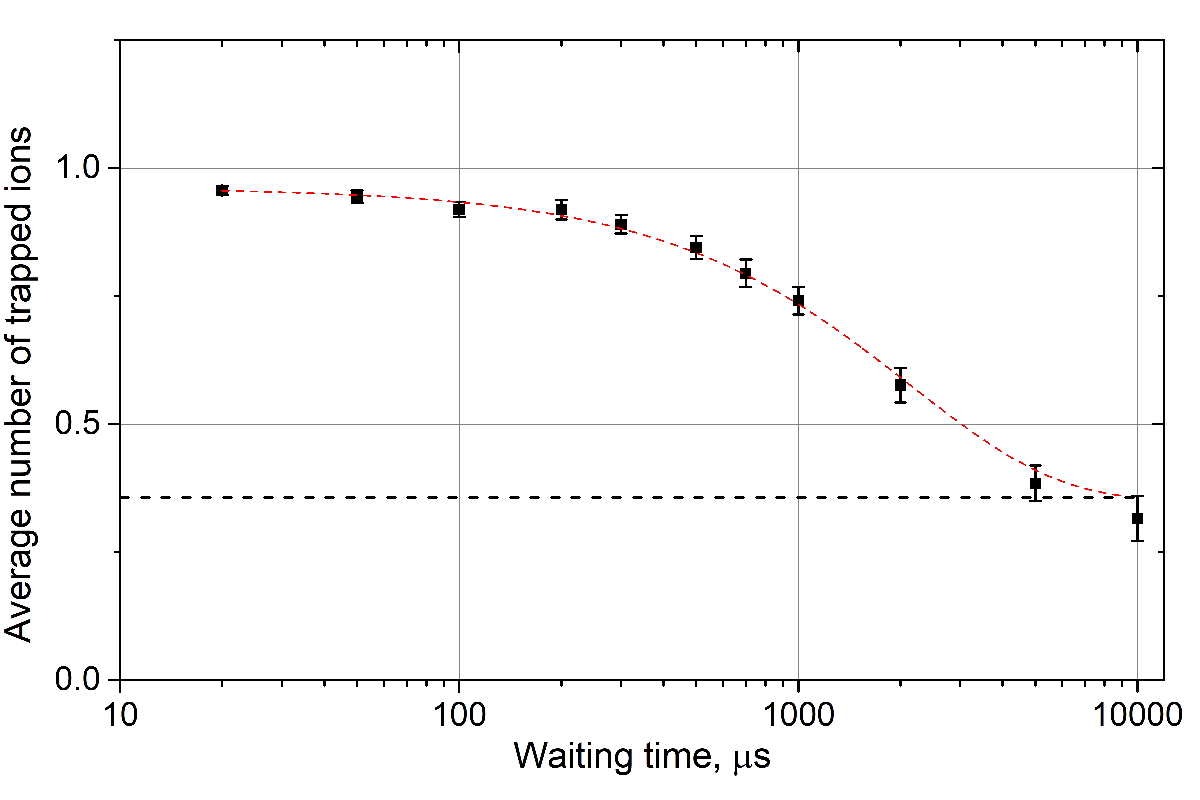}
    \caption{Average number of extracted Ca$^{+}$ ions as a function of waiting time. Loading time: 50\,\textmu s. Trapping field frequency $\Omega_\textrm{slow}$, voltage amplitude 35\,V. The red dashed line indicates the exponential decay. The horizontal dashed line indicates the background level when the ionization lasers are blocked.}
    \label{fig:ions_lifetime}
\end{figure}

To determine the stability of the orbits on which trapped particle move, we studied the evolution of the trapped particle number over a waiting period ranging from 10\,\textmu s up to 1\,s. Results for electrons are shown in Fig.\,\ref{fig:e-lifetime} and for ions in Fig.\,\ref{fig:ions_lifetime}. The initial number of extracted particles, $N_1$, depends on several parameters such as loading time and field amplitude as well as the overall quality of the trap which is affected, for example, by the alignment of the PCBs. Subsequently, the number drops exponentially with a decay constant $\tau$ until a plateau at $N_0$ is reached, as described by the fitting function,
\begin{equation}
    N = N_0 + N_1 e^{-t / \tau}\,. 
    \label{exp_decay}
\end{equation}
A similar behavior was previously reported with the same trap design in Berkeley and can be explained by a significant amount of electrons being produced near the fringes of the trap volume. Simulations show that, depending on the phase of the RF field at the time of ionization, these can reside on unstable orbits in the trap for an extended period~\cite{Matthiesen2021}. We estimate that losses due to collisions with background gas and atomic beam particles are negligible on the time scales studied here.

For both electrons and ions, we observe lifetimes of the initial population on the order of several milliseconds. The final population parameter $N_0$ includes contributions from trapped electrons and a small constant background due to spurious production of charged particles in the vacuum chamber. It is typically one electron per ten loadings 
as determined by blocking both laser beams while all other elements of the experiment, such as the atomic oven and the RF field, remain switched on. We find that often only one particle 
remains in the trap after 1\,s of wait time. The remaining particles appear to be on stable orbits. We attribute this to the small volume of the trap which in the ideal case has a harmonic potential only in the central 200 \textmu m.

\begin{figure}
    \centering
    \includegraphics[width=\linewidth]{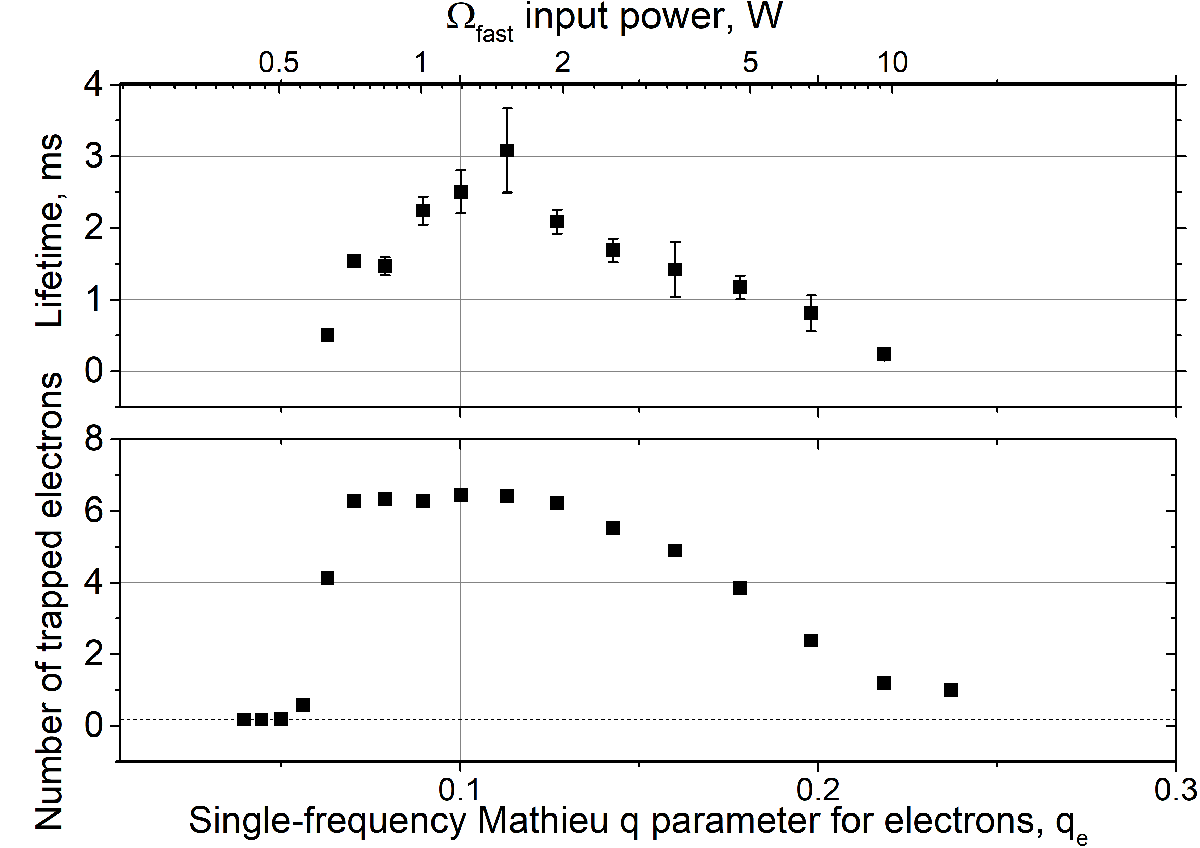}
    \caption{Trapping efficiency versus Mathieu $q_e$ parameter for electrons. Top panel: storage time of electrons, $\tau$; bottom: number of electrons, $N_0$, extracted after 50\,\textmu s waiting time. The dashed line indicates the background level.}
    \label{fig:e-lifetime VS Ghzpower}
\end{figure}

\begin{figure}
    \centering
    \includegraphics[width=\linewidth]{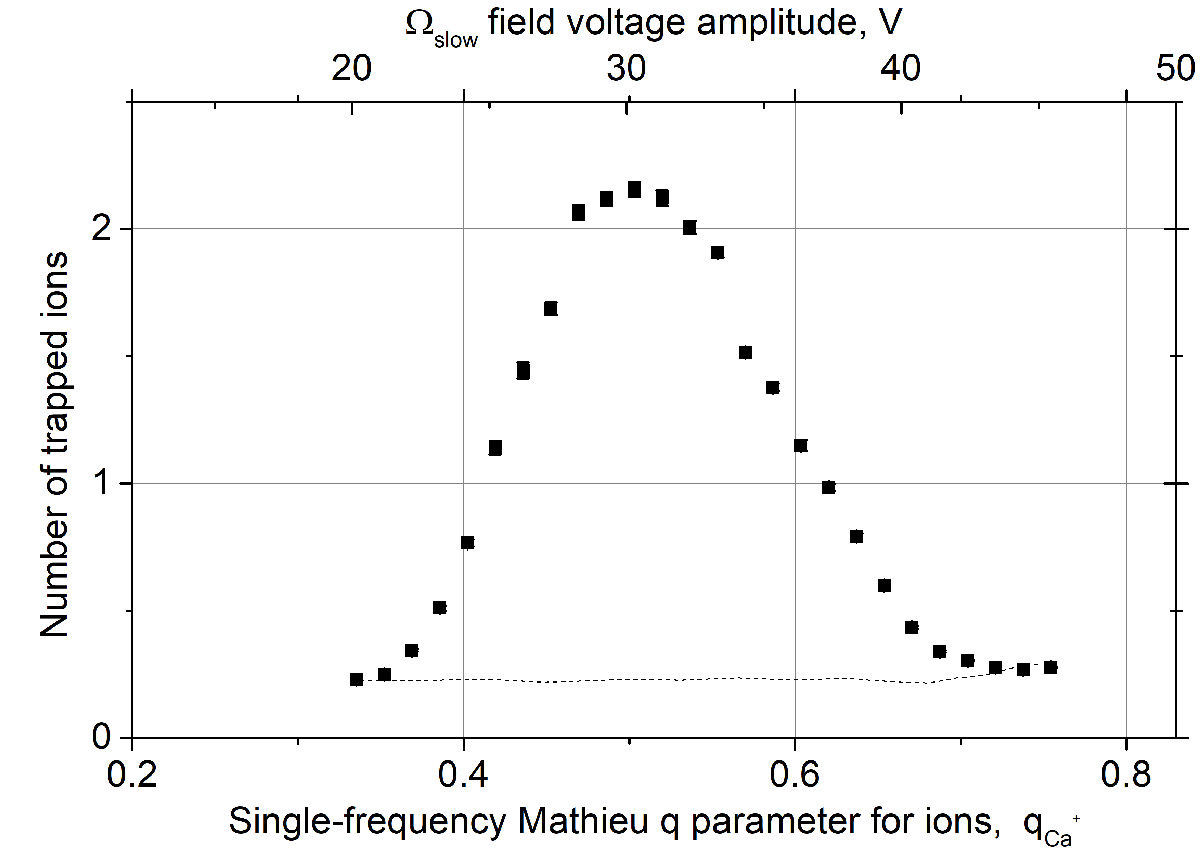}
    \caption{ Trapping efficiency versus Mathieu $q_{Ca^+}$ parameter for $Ca^+$ ions. Loading time: 100\,\textmu s, waiting time 50\,\textmu s. The dashed line indicates the background level.}
    \label{fig:ions_MHzampl}
\end{figure}

To investigate the region of stability for electrons, we conduct a series of measurements at various amplitudes of the $\Omega_\textrm{fast}$ field. We extract two characteristics to quantify the stability of particle orbits. First is the number of loaded electrons $N_0$ measured after a waiting time of 50\,\textmu s, and the second is the lifetime of electrons, see Fig.\,\ref{fig:e-lifetime VS Ghzpower}. Both measured characteristics show a clear dependence on the power of the $\Omega_\textrm{fast}$ field. The optimal value of $\Omega_\textrm{fast}$ input power for trapping is found to be approximately 1.2\,W which corresponds to a $q_e$ parameter of 0.11. Beyond this value, both the number and the lifetime of electrons 
drop. The lifetime curve has a sharper decline, which may indicate that at non-optimal values of the trap drive, electrons still can be trapped, but cannot be confined for a long time. The onset of instability at relatively low values of $q_e$ can be explained by the low mass of electrons, which leads to a larger motional amplitude than that of ions at the same value of $q$. A similar measurement was performed for ions in the $\Omega_\textrm{slow}$ field, where only the number of initially loaded ions was measured; results are shown in Fig.\,\ref{fig:ions_MHzampl}. Here, the onset of instability occurs for $q_{\textrm{Ca}^+}\gtrsim0.5$  as is typical for ion traps.

\subsection{Trapping electrons and ions with two frequencies applied}

\begin{figure}
    \begin{minipage}{\linewidth}
        \flushleft
        \includegraphics[width = \linewidth]{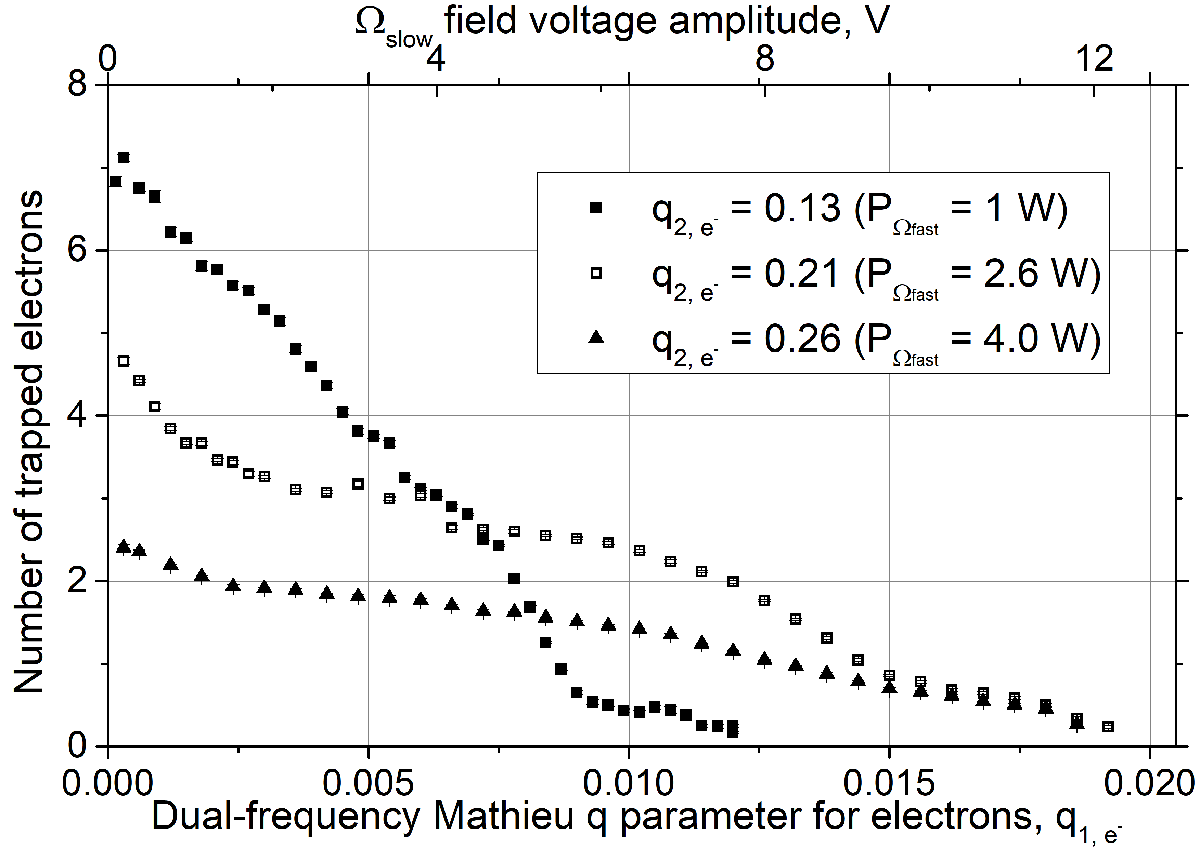 }
    \end{minipage}
    \begin{minipage}{\linewidth}
        \flushright
        \includegraphics[width = \linewidth]{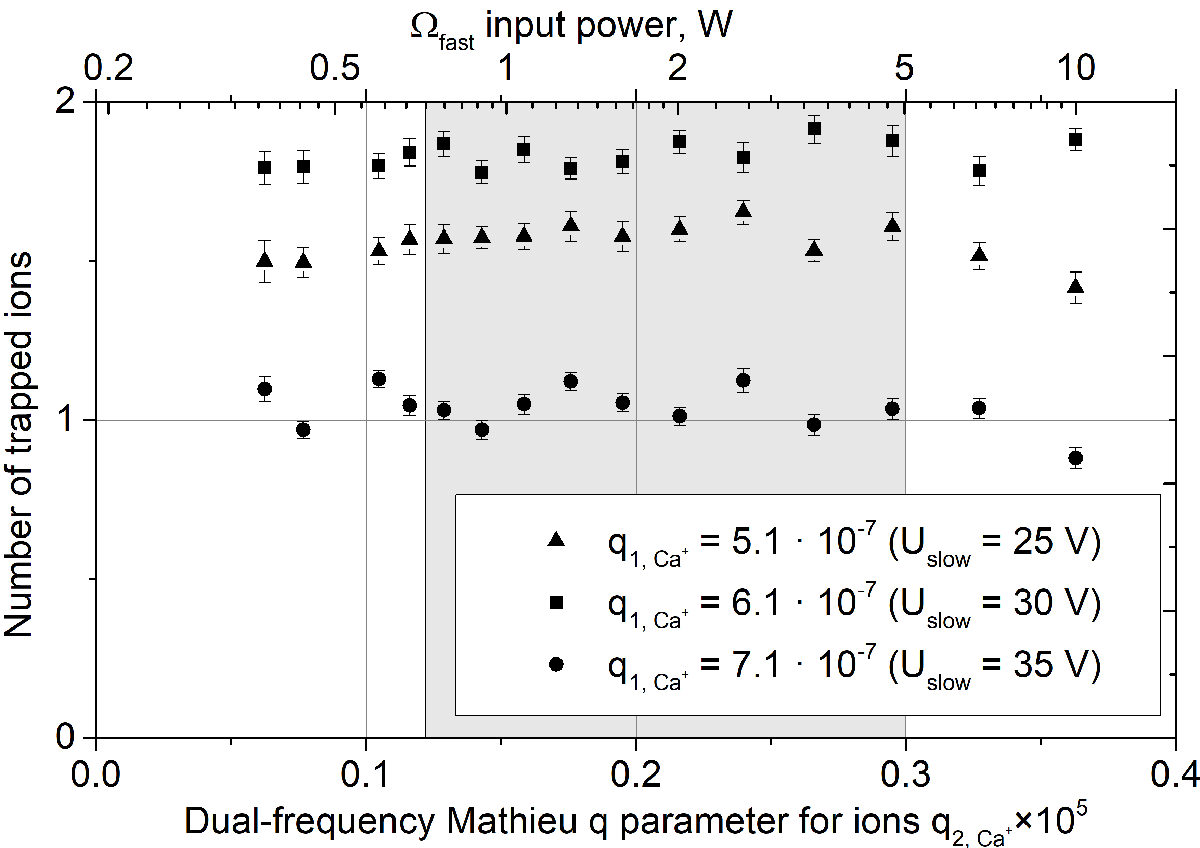}
    \end{minipage}
    \caption{Top: The dependence of the average number of trapped electrons on the amplitude of the $\Omega_\textrm{slow}$ voltage at a wait time (WT) of 50\,\textmu s. Bottom: the dependence of the average number of trapped Ca$^{+}$ ions on the power of $\Omega_\textrm{fast}$ field at WT = 50\,\textmu s and WT = 500\,\textmu s. The gray area indicates the range of the $\Omega_\textrm{fast}$ input power for stable electron trapping at a single frequency.}
    \label{fig:electrons&ionsl}
\end{figure}

Now that we know the optimal conditions for trapping electrons and ions in single RF fields and have ensured that the applied frequencies do not coincide with secular frequencies of the other species, we attempt to find conditions where both species can be trapped simultaneously. To this end, we trap electrons at different strengths of the $\Omega_\textrm{fast}$ field and investigate the trapping efficiency for electrons when applying the $\Omega_\textrm{slow}$ field for ion trapping on the ion trap electrodes. Similarly, we trap Ca$^+$ ions using different amplitudes of the $\Omega_\textrm{slow}$ field and investigate the trapping stability while driving the resonator at $\Omega_\textrm{fast}$. 
Results are shown in Fig.\,\ref{fig:electrons&ionsl}. Even small amplitudes of the $\Omega_\textrm{slow}$ field cause a significant decrease in the number of trapped electrons, and this value drops to zero at 12\,V of the $\Omega_\textrm{slow}$ voltage. Comparison to Fig.~\ref{fig:stability} shows that ideally $q_{2,\,e}$ should be increased to work in a region where larger amplitudes of the $\Omega_\textrm{slow}$ field can be allowed for stable trapping of electrons. But that requires the application of more power at $\Omega_\textrm{fast}$, which would damage our trap. Although theoretically stable co-trapping should be possible at the probed values of $q_{2,\,e}$, minor imperfections in the electrode shape and alignment seem to prevent this.

The situation for ions is completely different. We keep $q_{1,\,\textrm{Ca}^+}$ fixed and scan $q_{2,\,\textrm{Ca}^+}$ from $0.06\times 10^{-5}$ to $0.4\times 10^{-5}$. We find that even high voltages of the $\Omega_\textrm{fast}$ field do not affect the number of trapped ions. The different response to the second frequency field can be explained by the mass difference between electrons and ions. While the fast oscillations of the $\Omega_\textrm{fast}$ field average to zero on the time scales of the slow motion of the ions, the slower-oscillating trapping field for ions corresponds to a quasi-DC potential for electrons.

\section{Conclusions and Outlook}

In this work, we have demonstrated the suitability of our trap geometry for storing either electrons or ions in the same trap volume while applying two RF fields simultaneously. When operating at a single frequency, we observe that tens of electrons or ions can be stored with a storage time of several milliseconds. Typically, one electron or ion remains in the trap for at least 1\,s. In dual-frequency mode, the measured number of trapped ions does not depend on the amplitude of the $\Omega_\textrm{fast}$ field, while the number of trapped electrons decreases by approximately 10\,\% for every 1\,V increase in the $\Omega_\textrm{slow}$ voltage. 
This behavior is in agreement with our theory where we found that, at the accessible $\Omega_\textrm{fast}$ power, stable co-trapping is sensitive to imperfections in the $\Omega_\textrm{slow}$ field which is exacerbated in our non-orthogonal electrode geometry. This can also be explained by differences in the particle masses and corresponding oscillation frequencies and amplitudes, showcasing the difficulties with trapping electrons in an RF trap.

To address the issues that prevent us from co-trapping electrons and ions, we are developing a next-generation trap. Misalignment of electrodes due to mechanical tolerances and thermal cycling when applying high RF power distorts the quadrupole field. Furthermore, the surface roughness and sharp edges of PCB electrodes introduce local electric field cusps that further distort the quadrupole field so that the trap volume is reduced and heating rates are increased~\cite{Lin2016, Ou2016}. Our theory predictions show that the region of stability can be increased by implementing orthogonal electrodes for the MHz field. Additionally, we aim to reduce the bare dielectric surface area of the PCBs exposed to photoelectrons and ions which tend to accumulate static charge thereby perturbing the trapping potential noticeably on a few-hours timescale. To address these challenges, we plan to fabricate a new trap from a more rigid material which allows the precision and stability of trap-element alignment to be improved.
We also plan to minimize the exposure of dielectric surfaces by ensuring they are electrostatically shielded from the trap center, and to fabricate electrodes with smoother surfaces and gentler curvatures. All these requirements can be met with the selective-laser-etching fabrication technology used in state-of-the-art ion traps \cite{SLEtrap, Romaszko2020285}.

In future work we plan on trapping cold positrons generated in $\beta^{+}$ decay of $^{22}$Na that are moderated and cooled in a buffer-gas trap \cite{Surko2015}. Co-trapping these with normal matter will allow us to pursue a rewarding scientific program including the investigation of bound positron-atom systems \cite{Harabati2014, Gribakin2015}. Moreover, our experiments pave the way for efficiently generating antihydrogen from its elementary constituents by storing both positrons and antiprotons in the same RF trap volume. In this regard, the results presented here are promising as they suggest that co-trapping of antiprotons, which have a lower mass than $^{40}$Ca$^+$ ions, and positrons in RF traps seems to be achievable by careful adjustment of the lower-frequency field. This method is attractive because it allows arbitrarily long interaction times between positrons, antiprotons, and normal matter.

\noindent
\begin{acknowledgments}
The authors thank Masaki Hori for his important contributions to the design and implementation of the experiment, Nathan Leefer for helpful discussions, Kai Krimmel for his contribution at the early stages of the experiment, Lei Cong for his help with the theoretical calculations, and Ron Folman for his original proposal of 'antimatter on a chip'. Michal Hejduk thanks the Czech Science Foundation (GA\v{C}R: GA24-10992S) for the support. Niklas Lausti thanks the Charles University Grant Agency (GAUK 131224), and the Czech Ministry of
Education, Youth, and Sports (project QM4ST, id. no. EH22 008/0004572) for the support.The study was supported in part by the DFG Project ID 390831469: EXC 2118 (PRISMA+ Cluster of Excellence), the DFG project TACTICa ID 495729045, and the project ``Quantum Sensing for Fundamental Physics (QS4Physics)’’ funded by the Innovation pool of the
research field Helmholtz Matter of the Helmholtz Association.
\end{acknowledgments}

\appendix
\section{EMT calibration}
\label{sec:emt-calibration}
\begin{figure}[h!]
    \centering
    \includegraphics[width=\linewidth]{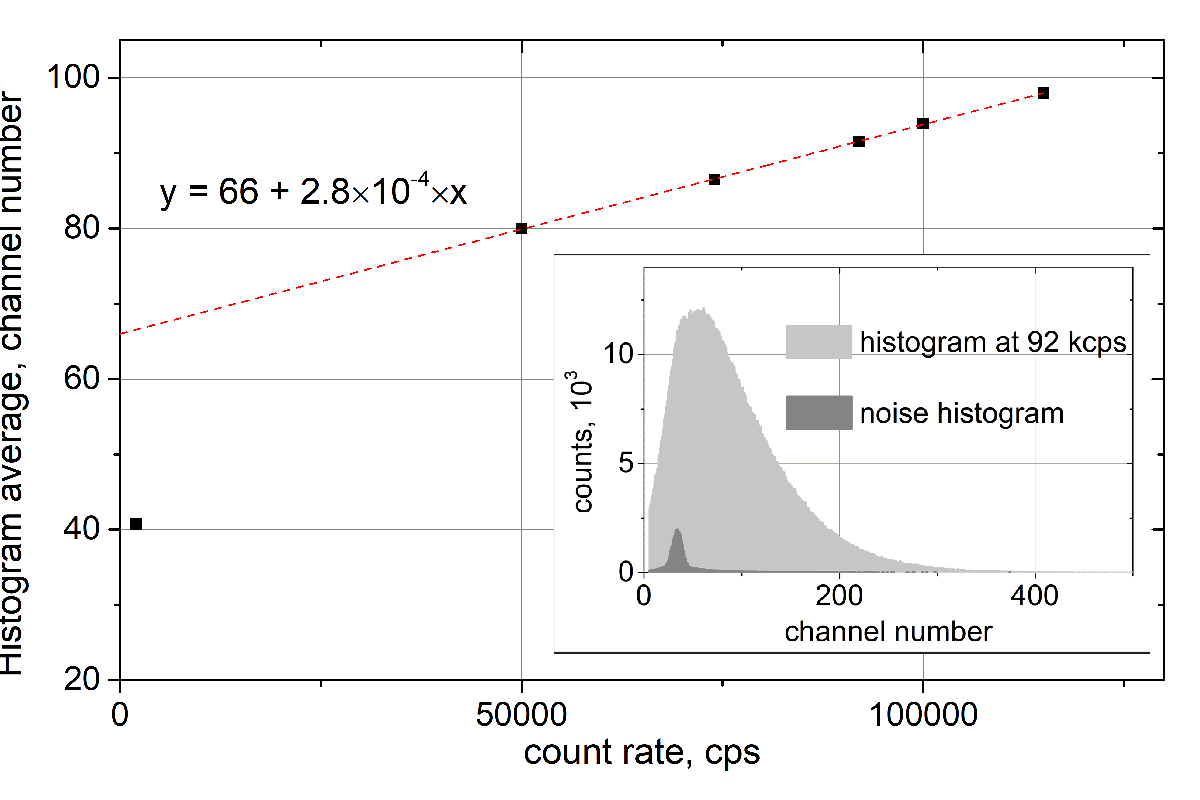}
    \caption{Calibration of the detection chain in case of electron detection. (black) -- histogram averages after $10^{6}$ repetitions, (red dashed) -- linear approximation. Inset: a histogram of the EMT+CSP readout at 92\,cps count rate and a histogram of detector noise, acquisition time 30\,s}
    \label{fig:emt_calibration}
\end{figure}

As described in Section~\ref{sec:setup}, the detection chain includes a Hamamatsu R596 electron multiplier tube (EMT), an Ortec 142PC charge-sensitive preamplifier (CSP), and a CAEN 5770 digitizer. Each primary electron or ion entering the EMT produces in the order of $10^{6}$ electrons at the output. The resulting charge pulse is amplified by the CSP and then counted as a detection event of which the voltage amplitude is measured using the 14-bit digitizer. EMT pulses resulting from single-electron events exhibit a Poisson-like amplitude distribution due to the statistical nature of secondary-electron emission~\cite{Saldanha2016}. We denote this single-electron response by $P_1$. If two electrons arrive at the EMT simultaneously, or within the pile-up time of the digitizer, they are counted as a single event and produce a pulse with double amplitude distribution $P_2$. When the number of electrons per detection event varies during the acquisition time, the resulting pulse amplitude histogram consists of multiple overlapping distributions: $P_n = P_1 + P_2 + \ldots$  as shown in the inset of Fig.~\ref{fig:emt_calibration}. The average number of electrons per detection event can then be calculated as: $N = \bar{P}_n / \bar{P}_1$

To determine $\bar{P}_1$, we employ a simple model-independent approach. The digitizer is switched to continuous detection mode with internal triggering. Electrons are continuously produced by photoionization, whereas the quadrupole trapping field and extraction pulses are turned off. The typical EMT and holder potentials are applied so that most of the electrons produced hit the first dynode. We assume that the arrival of the electron in the EMT follows a Poisson distribution with an average interval of $1 / cps$, with $cps$ the number of detection events per second. At a specific range of count rate values, the experimental distribution $P$ can be written as a linear function on $cps$: $P(cps) = P_b + P_1 + a P_2 \cdot cps$, where $P_b$ is the background noise and $a$ is the weight of two-electron events. Then $\bar{P}(cps) = \bar{P}_b+\bar{P}_1+a\bar{P}_2\cdot cps$. By subtracting noise and extrapolating $\bar{P}(cps)$ from the linear region to the zero count rate, we find the $\bar{P}_1$. We vary the count rate from 50 to 115\,kcps and determine the single-electron peak average $\bar{P}_1$ to be in channel $66\pm0.2$. A similar approach is applied for the ions. The main source of uncertainty in this approach is the difference between the measured detection event count rate and the true electron count rate. According to our estimation, this systematic error can be as high as 20\%.

\bibliography{references.bib}%

\end{document}